\documentclass[aps,prd,amssymb,cite,
amsfonts,epsf,preprintnumbers,nofootinbib,superscriptaddress]{revtex4}
\usepackage[dvips]{graphicx}
\usepackage{bm,latexsym,amsmath,amssymb,amsfonts}
\usepackage[usenames,dvipsnames]{color}
\usepackage[colorlinks=true,linkcolor=blue]{hyperref}
\usepackage{color}
\usepackage{ulem}
\usepackage{epsfig}
\usepackage{braket}
\usepackage[mathscr]{eucal}
\usepackage{cancel}
\usepackage{mathrsfs}
\usepackage{pgf,tikz}
\usepackage{slashed}
\usetikzlibrary{arrows,automata}
\definecolor{mypink1}{rgb}{0.858, 0.188, 0.478}
\definecolor{mypink2}{RGB}{219, 48, 122}
\definecolor{mypink3}{cmyk}{0, 0.7808, 0.4429, 0.1412}
\definecolor{mygray}{gray}{0.6}
\definecolor{pptbg}{rgb}{0.961,0.945,0.863}

\newcommand{\be}[1]{\begin{equation} \label{#1}}
\newcommand{\ee}{\end{equation}}
\newcommand{\bea}{\begin{eqnarray}}
\newcommand{\eea}{\end{eqnarray}}
\newcommand{\ba}{\begin{array}}
\newcommand{\ea}{\end{array}}

\newcommand{\bel}{\begin{align}}
\newcommand{\eel}{\end{align}}

\newcommand{\cT}{\mathcal{T}}

\newcommand{\hreff}[1]{\href{#1}{\color{blue}{#1}} }

\begin{document}
\title{Temperature upper bound of an ideal gas}

% more complex case: 4 authors, 3 institutions, 2 footnotes
\author{Hyeong-Chan Kim}
\affiliation{School of Liberal Arts and Sciences, Korea National University of Transportation, Chungju 380-702, Korea}

% e-mail addresses: one for each author, in the same order as the authors
\email{hckim@ut.ac.kr}

%% simple case: 2 authors, same institution
%% \author{A. Uthor}
%% \author{and A. Nother Author}
%% \affiliation{Institution,\\Address, Country}

% more complex case: 4 authors, 3 institutions, 2 footnotes
%\author[a]{Hyeong-Chan Kim}
%\affiliation[a]{School of Liberal Arts and Sciences, Korea National University of Transportation, Chungju 380-702, Korea}

% e-mail addresses: one for each author, in the same order as the authors
%\emailAdd{hyeongchan@gmail.com}
\begin{abstract}
We study thermodynamics of a heat-conducting ideal gas system.
The study is based on i) the first law of thermodynamics from action formulation which expects heat-dependence of energy density and ii) the existence condition of a (local) Lorentz boost between an Eckart observer and a Landau-Lifschitz observer--a condition that extends the stability criterion of thermal equilibrium.
The implications of these conditions include:
i) Heat contributes to the energy density through the combination $q/n\Theta^2$ where $q$, $n$, and $\Theta$ represent heat, the number density, and the temperature, respectively. 
ii) The energy density has a unique minimum at $q=0$. 
iii) The temperature upper bound suppresses the heat dependence of the energy density inverse quadratically. 
This result explains why the expected heat dependence is difficult to observe in ordinary situation thermodynamics.
\end{abstract}

\maketitle
%\flush bottom

The lower bound of temperature, referred to as the absolute zero in Kelvin, is determined by the behavior of the ideal gas volume (or pressure) in thermal equilibrium, contingent upon the temperature changes. 
On the other hand, a precise upper bound of temperature lacks a unique definition. 
In the theory of particle physics, the `Hagedorn temperature'~\cite{Hagedorn} for hadrons serves as an upper bound. 
Beyond this temperature ordinary matter is no longer stable, and must either ``evaporate" or convert into other (quark-gluon) phase.
Similarly,  the string Hagedorn temperature~\cite{Giddings:1989xe} plays a comparable role for strings.
Given that temperature is a thermodynamic quantity, it is natural to inquire about the implications of the existence of the temperature upper bound on thermodynamics.
In this work, we show that it affects the nature of heat conduction.

In the kinetic theory picture, an ideal gas is a theoretical gas consisting of numerous randomly moving, non-interacting point particles. 
The model is useful as it adheres a simplified equation of state known as the ideal gas law, effectively approximating the states of most physical gases under non-extreme conditions.
The thermodynamic properties of an ideal gas can be described by two equations.
The first is the ideal gas law, $ \Psi V=N k_B \Theta$, where $\Psi$, $V$,  $N$, $k_B$ and $\Theta$ denote the pressure, the volume, the total number of particles, the Boltzmann constant, and the temperature, respectively.
Dividing both sides of the equation by the volume $V$ and adopting the natural units with $k_B=1$, the law presents a simple relation between pressure and temperature:
\be{idgas1}
\Psi = n \Theta,
\ee
where $n \,(\equiv N/V)$ denotes the number density.
 Because we are interested in the relativistic thermodynamics based on local description of the theory, we write physical quantities in terms of densities such as $n$.  
The other equation expresses the energy density $\rho \,(\equiv E/V)$ of the gas consisting of particles with mass $m$ as:  
\be{rho:ig}
\rho(n,s) =nm + c_v n \Theta, 
\ee
where $c_v$ is a constant denoting the dimensionless specific heat capacity at constant volume, approximately $3/2$ or $5/3$ for monoatomic or diatomic gases, respectively.
Other quantities can be obtained from these two laws. 
For example, one can calculate the entropy using the first law of thermodynamics.
Then, the specific entropy, $\sigma^* \equiv  s^*/n$, where $s^*\equiv S^*/V$ denotes the entropy per unit volume for the system in thermal equilibrium\footnote{Here, the asterisk denotes a thermal equilibrium quantity. For example, the character ${S}^*$ represents the entropy in thermal equilibrium. The character without the asterisk, $S$, is designated to represent that with heat flux.}, takes the form 
\be{sigma*}
\sigma^* = c_v \log \frac{\Theta}{\Theta_0} - \log \frac{n}{n_0} .
\ee
This ideal description of the gas is valid only when the following assumptions hold:
First, the atoms or molecules do not interacts or collides perfectly elastically with each other, so one can ignore intermolecular attractions.
Second, the particles are so small that one can ignore their volume relative to the volume occupied by the gas.
In a general situation such as at ultra-high temperatures, the pressure of the gas may not be written in this form due to an increase in collisions between particles, but follows different rules, e.g., in polytropic form $\Psi  \propto \rho^\gamma$. 
However, for gases satisfying the ideal gas assumptions, it has been shown that the pressure still follows the rule~\eqref{idgas1}, even for ultra-high temperatures or in the presence of a strong gravity~\cite{Kim:2016xnt}, with slightly modified specific heats.
Because we are considering a generalization of an ideal gas, we still adhere to the law~\eqref{idgas1}.
When dissipation occurs between gases or when the system size is too small to experience intermolecular forces, the ideal gas approximation may not be valid. 
This condition necessitates that the system size be larger than the atomic scale.

Given the utility of the ideal gas approximation, several attempts have been made to generalize the ideal gas.
Generalizations of the ideal gas obeying fractional statistics were  explored~\cite{Wu1994}.
In the context of self-gravitating systems, the ideal gas under the influence of Newtonian~\cite{Kim:2016txr} and Einstein gravity~\cite{Louis-Martinez:2010jvm,Kim:2016xnt} were investigated.
These studies were based on the energy density derived from  thermal equilibrium systems.
References~\cite{Holyst2022,Holyst2023} demonstrated the possibility of describing out-of-equilibrium system in stationary state using the ideal gas, portraying global thermodynamic functions.
In their works, the authors developed a heat conducting massive mono-atomic ideal gas model.
They considered $N$-gas particles of mass $m> 0$ contained in a square box maintaining local thermal equilibrium. 
When the temperatures at both ends of the box are $\Theta_2$ and $\Theta_1$, respectively, and heat flows continuously,
the total entropy of the system is\footnote{In their work, they considered a monoatomic gas, so $c_v = 3/2$.}, 
\begin{equation}\label{S:Holyst}
S = S^* + N(c_v+1) \log \left[ \left(\frac{\Theta_2}{\Theta_1}\right)^{1/2} 
	\frac{\log (\Theta_2/\Theta_1)}{\Theta_2/\Theta_1 -1} \right]
\approx S^* - \frac{N (c_v+1)}{24} \left(\frac{\Theta_2}{\Theta_1} -1\right)^2 +\cdots ,
\end{equation}
where in the second equality we use the approximation $\Theta_2 /\Theta_1 \sim 1$.
Here, $S^*$ is interpreted as $\sigma^*V$ from Eq.~\eqref{sigma*} for an equilibrium state with the same total energy and volume as the heat-conducting model. 
Because both the total entropy $S$ of the heat-flowing system and the entropy $S^*$ of the corresponding equilibrium system are extensive quantities, their difference $\Delta S \equiv S-S^*$ is also an extensive quantity.
When $\Theta_2 \to \Theta_1$, we easily notice $\Delta S  \to 0$.
However, when $\Theta_2 \neq \Theta_1$, heat flows, and the entropy becomes dependent on the temperature difference (and consequently on heat).

Noting that relativistic description requires local forms for physical quantities, we need to denote the contribution of heat in a local form. 
To accomplish this purpose, we consider a macroscopic system of thickness $L$ consisting of several subsystem layers of thickness $\delta L$ along $x$-direction, which have the form described above and satisfy $\sum \delta L = L$ with $\delta L \ll L$.
Now, we assume that $\delta L$ is small enough on a macroscopic scale, even though it should be still large enough on a microscopic scale so that a statistically sufficient number of particles are contained in the box, atomic interactions can be ignored, and quantum properties of the particles are hidden.  
Let $\Theta_2 = \Theta(x+ \delta L)$ and $\Theta_1 = \Theta(x)$ denote the temperatures of the ends of a subsystem, respectively.
Then, $\Theta_2 \approx \Theta_1$ and we can use the approximation, 
$$
\frac{\Theta_2}{\Theta_1}-1 \approx \delta L \frac{\partial_x \Theta} {\Theta} = -\delta L \frac{q}{\kappa \Theta},
$$ 
where we use the Newtonian definition of heat, $q_k = -\kappa \nabla_k \Theta$.
 Here, $\kappa$ denotes the heat conductivity.
Even though it vanishes in the $\delta L \to 0$ limit, we keep the first order term to express the contribution of heat. 
This assumption is justified because one cannot ignore this value $\delta L$ on a microscopic scale.
Later in this work, we evaluate the heat conductivity of an ideal gas and write this formula in a form proportional to the ratio between this length $\delta L$ and the light traveling distance during the mean free time of the gas particles. 
Dividing both sides of Eq.~\eqref{S:Holyst} by $N$, the difference in specific entropy, $\delta \sigma \equiv \sigma -\sigma^*$, has a non-trivial value:
\be{delta sigma}
\delta \sigma \equiv \sigma - \sigma^* \approx  - \frac{c_v+1} {24}\left(\frac{\delta L\,  q}{\kappa \Theta}\right)^2 .  
\ee
From now on, the notations such as $\Theta(x^a)$, $\rho(x^a)$, $n(x^a)$, $s(x^a)$, $\sigma(x^a)$ and $q(x^a)$ denote the thermodynamic quantities defined at the subsystem located at the point $x^a$. 
The `locality' we mention is defined in this way. 
%as local quantities in the sense that they denote those of a subsystem at the point $x^a$.   

In the realm of relativistic thermodynamics, the variational formulation~\cite{Taub54,Carter72,Carter73,Carter89,Andersson:2013jga,AnderssonNew,LK2022,Kim:2023lta} suggests that the energy density $\rho$ should depend not only on the typical number and entropy densities but also on heat.
Such a dependence does not manifest in the original ideal gas model in Eq.~\eqref{sigma*}, or even for other well-known equation of states constructed based on thermal equilibrium state. 
In this letter, we explore the incorporation of heat dependence such as in Eq.~\eqref{delta sigma} into an ideal gas equation of state without compromising the assumptions inherent in ideal gas behavior. 
There are several compelling reasons to consider incorporating heat dependence into the energy density. 
Notably, many (astro-)physical objects exhibit heat conduction. 
The investigation of such out-of-equilibrium systems relies heavily on an understanding of states involved in heat conduction. 
 The connection between gravity and thermodynamics~\cite{Bekenstein:1974ax,Jacobson:1995ab,Padmanabhan:2009vy,Verlinde:2010hp,Carlip:2014pma,Cocke,Kim:2019ygw,Lee:2008vn,Lee:2010bg}, highlighted since Hawking's work on black hole thermodynamics~\cite{Hawking:1974sw}, further motivates investigations into gravitating systems.
Examining self-gravitating systems in thermal equilibrium has been a longstanding effort, contributing significantly to our comprehension of astrophysical systems. 
Notably, studies on the entropy of spherically symmetric self-gravitating radiation and its stability have been conducted over the years~\cite{Sorkin:1981,Gao:2016trd,Roupas:2014nea,Kim:2019ygw}. 
These investigations have demonstrated that the requirement of maximum entropy for self-gravitating radiation in a spherical box reproduces the Tolman-Oppenheimer-Volkoff equation for hydrostatic equilibrium~\cite{Oppenheimer,Tolman3,Cocke}.
The necessity for studying out-of-thermal equilibrium situations in these areas is high, which requires an understanding of heat dependence.
We endeavors to establish a foundation for future research in understanding heat dependence in out-of-equilibrium situations in this work.

\vspace{.2cm} 
The variational formulation for relativistic thermodynamics involves a Lagrangian-like function, often referred to as the master function \(\Lambda\), which incorporates matter and entropy fluxes. 
At the second order in the deviation from thermal equilibrium~\cite{Priou1991}, this formulation is essentially equivalent to the Israle-Stewart theory~\cite{IS1,IS2,IS3}, known to be stable and causal~\cite{Hiscock1983}.
The variational formulation has advantage of naturally leading to an extended Gibbs relations in various models of extended irreversible thermodynamics. 
The aximatic formulation has been extended to encompass dissipations and particle creations~\cite{Andersson:2013jga,AnderssonNew}. 
Recently, the binormal equilibrium condition was proposed~\cite{LK2022} to address the perceived incompleteness~\cite{Andersson2011} of relativistic heat conduction theory.
The steady thermal state were studied based on this proposal~\cite{Kim:2023lta}.

In this letter, we consider a two-fluid system consisting of a number flux \(n^a\) and an entropy flux \(s^a\).
Typically, the heat conduction equation is expressed in the Eckart decomposition, where the observer's four-velocity \(u^a\) is parallel to the number flux.
The misalignment between the entropy flux \(s^a\) and the number flux \(n^a\) gives rise to entropy creation generated by the heat flux \(q^a\). 
 Explicitly, given the number density \(n\), entropy density \(s\), and the heat flux \(q^a\), the particle number and the entropy fluxes are
\be{na}
n^a \equiv n u^a, \qquad s^a \equiv s u^a + \varsigma^a; \qquad \varsigma^a \equiv \frac{q^a}{\Theta},
\ee
where the heat flux $q^a$ is normal to the matter flow, \(q^a u_a = 0\).
Then, the heat strength is defined to be $q \equiv \sqrt{q^a q_a}$. 
This procedure uniquely defines heat in a coordinates-independent manner at least for this two-fluid model. 
Assuming that matter does not have explicit directional dependence, the fluid Lagrangian must be a function of scalars composed of the matter and caloric fluids.  
Therefore, $\Lambda$ depends on the three scalars: $n \equiv \sqrt{-n^a n_a}$, $s \equiv -u^a s_a$, and $\varsigma  \equiv \sqrt{\varsigma^a \varsigma_a}$, constructed from $n^a$ and $s^a$.
Starting from the master function \(\Lambda(n,s,\varsigma)\), the energy density was constructed through the Legendr\'{e} transformation: \(\rho(n, s, \vartheta) = \varsigma \vartheta - \Lambda\)~\cite{Carter89}, where $\vartheta$ is the conjugate variable to $\varsigma$. 

In this work, we use two distinct results of the relativistic thermodynamics.
First is the first law of thermodynamics (the generalized Gibbs relation)~\cite{Carter89,AnderssonNew}: 
\be{1st law}
d\rho(n,s,\vartheta) = \chi dn + \Theta ds + \varsigma d\vartheta,
\ee
derived from the variational law of the master function, 
where \(\chi\) denotes the chemical potential.
The extended Gibbs relation resembles those postulated in many approaches to extended thermodynamics~\cite{Jou1993}.
A key distinction is that this relation emerges naturally from the variational relation of the action.
Note that the variation of the specific entropy~\eqref{delta sigma} contains the heat dependence naturally, which signifies the appearance of the last term of the extended Gibbs relation~\eqref{1st law}.
Noting the result, it is natural to ask when does this heat dependence appears and what is its role in thermodynamics.

%\vspace{3cm}
The second is the stability condition of a thermal equilibrium state, derived from perturbative studies based on relativistic thermodynamics. 
Even when the geometry is dynamical, (local) thermal equilibrium with respect to a comoving observer is characterized by the vanishing of the Tolman vector, $ \cT_a \equiv \frac{d}{d\tau} (\Theta u_a) + \nabla_a \Theta=0,$ where \(d/d\tau \equiv u^b \nabla_b\) and \(\nabla_a\) denotes the covariant derivative for a given geometry. 
When the spacetime is static, the Tolman temperature gradient~\cite{Tolman,Tolman2},
$
\Theta(x^i) = T_\infty/\sqrt{-g_{00}(x^i)},
$
appears naturally from the equation.
Here \(g_{00}\) and \(T_\infty\) represent the time-time component of the metric on the static geometry and the physical temperature at the zero gravitational potential hypersurface usually located at spatial infinity.
There were arguments~\cite{Lima:2019brf,Kim:2021kou} for the modification of the original form of the temperature; however, the result is generally accepted because of the universality of gravity~\cite{Santiago:2018kds,Santiago:2018lcy} and the maximum entropy principle~\cite{Sorkin:1981,Gao:2016trd}. 
For a two-fluid system with one number flow \(n^a\), the other equation characterizing thermal equilibrium is Klein's relation~\cite{Klein49,Kim:2021kou}.
The stability and causality of the thermal equilibrium state were also analyzed~\cite{Hiscock1987,Olson1990,LK2022}. 
For the thermal equilibrium state to be stable, the thermodynamic quantities must satisfy the inequality\footnote{Originally, this equation was stated in the form $\frac1n \left[1- \frac{\rho +\Psi}{\Theta} \beta \right] < 0$ for a thermal equilibrium, where $\vartheta = \beta q$ by definition and the regularity of $\beta$ is implicitly assumed. Originally this inequality was derived perturbatively as a stability condition of thermal equilibrium. Because the result was proven even when $q \neq 0$ non-perturbatively~\cite{Kim:2023lta}, we use the heat $q$ explicitly in the formula.} 
\be{stability:TTES}
\frac1n \left[1- \frac{\rho +\Psi}{\Theta} \frac{\vartheta}{q} \right] < 0,
\ee
where  \(\Psi = n \chi+s\Theta-\rho\) denotes the pressure of the system.
Recently, this inequality was re-derived~\cite{Kim:2023lta} as an {\it existence condition for Local Lorentz Boost} based on the fact that there exists a local Lorentz transformation between the Eckart~\cite{Eckart:1940aa} and the Landau-Lifshitz observers~\cite{LL} for a generic values of $q\neq0$ because the comoving vectors of both observers are time-like. 
Because the proof is non-perturbative, the inequality~\eqref{stability:TTES} holds even for out-of-equilibrium states.
This result presents a crucial observation on the stability and temperature of a thermodynamic system with heat conduction, as will be shown soon.
This inequality constrains the value of \(\vartheta\) to be larger than \(\Theta q/(\rho+\Psi)\).  
Therefore, this inequality must be a truth rather than a stability condition unless there does not exist a local Landau-Lifschitz decomposition.

\vspace{.2cm}

As shown in Eq.~\eqref{1st law}, the first law of thermodynamics, derived from the variational formulation, naturally predicts that the energy density $\rho$ will have contributions from heat. 
Therefore, we anticipate that the energy density is expressed as a function of three quantities: $n$, $s$, and $\vartheta$, where $\vartheta$ contains the heat dependence. 
Let us explore how heat contributes to the density when the system is out-of-equilibrium. 
When the interactions between molecules are activated, such as collisions between molecules, heat will propagate in a different direction from that of the number flux allowing $q^a \neq 0$ in Eq.~\eqref{na}. 
Since we aim for the system to be described by the ideal gas law, we require that the energy density $\rho = \rho(n, s, \vartheta)$ formally retains the same structure as that without heat contribution, as depicted in Eq.~\eqref{rho:ig}.
Then, the dependence of $\rho$ on $\vartheta$ can be inferred through the temperature, $\Theta = \Theta(n,s,\vartheta)$. 
Because we develop the model based on the thermal equilibrium case, the model reproduces the equilibrium result when $q=0$.
The reproduction of ordinary thermal equilibrium, based on the instantaneous velocity distribution, is evident in the $\Theta_2 =\Theta_1$ limit of Eq.~\eqref{S:Holyst}.

Now, we additionally require that both the ideal gas law~\eqref{idgas1} and the constancy of the specific heat capacity for constant volume are also kept.
We constrain the form of the energy density by implementing these conditions one by one.
From the first law, using the assumption that $c_v$ is constant, the temperature and the chemical potential become
\be{Tchi}
\Theta(n,s,\vartheta) \equiv
\left(\frac{\partial \rho}{\partial s}\right)_{n, \vartheta} = c_v n \left(\frac{\partial \Theta}{\partial s}\right)_{n, \vartheta} , \qquad
\chi(n,s,\vartheta) = 
\left(\frac{\partial \rho}{\partial n}\right)_{s, \vartheta} 
= m+ c_v \Theta + c_v n \left(\frac{\partial \Theta}{\partial n}\right)_{s, \vartheta} .
\ee
From the ideal gas law~\eqref{idgas1} and the thermodynamic identity $\rho +\Psi = n\chi+ s \Theta$, we get 
$
\chi = m + \Theta(c_v +1 -\sigma), 
$
where $\sigma \equiv s/n$ denotes the specific entropy, entropy per particle.
Comparing this equation with the last equation of Eq.~\eqref{Tchi}, we get 
\be{dT:n}
\frac{n}{\Theta} \left(\frac{\partial \Theta}{\partial n}\right)_{s, \vartheta} 
	= \frac{1-\sigma}{c_v}.
\ee
Solving the first equation in Eq.~\eqref{Tchi} and Eq.~\eqref{dT:n}, we get the temperature formally
\be{T:gen}
\Theta(n,s,\vartheta) = \Theta_0 \left(\frac{n}{n_0}\right)^{1/c_v}  \exp\left[\Phi(\vartheta) +\frac{\sigma}{c_v} \right],
\ee
where $\Phi(\vartheta) $ is an unidentified function of $\vartheta$ only independent of $n$ and $s$, which will define how the density varies with heat.
When $\Phi(\vartheta) =0$, this formula reproduces that of the thermal equilibrium.
Here, we check the constancy of the specific heat capacity for constant volume and $\vartheta$: 
\be{cv}
 c_{v,\vartheta} \equiv \Theta \left(\frac{\partial \sigma}{\partial \Theta}\right)_{n,\vartheta} = c_v .
\ee
The first law~\eqref{1st law} also presents $\varsigma$ from the variation of the energy density with respect to $\vartheta$:
$$
\frac{q}{\Theta} = \varsigma \equiv  \left(\frac{\partial \rho}{\partial \vartheta}\right)_{n,s} 
	= n c_v \Theta \Phi'(\vartheta) .
$$
Rewriting this equation, we get a relation between $\vartheta$ and the heat $q$:
\be{f'}
 \Phi'(\vartheta)  \equiv \frac{d \Phi(\vartheta) }{d\vartheta}= \frac{q}{c_v n\Theta^2}.
\ee

Until now, we have not fixed the form of the function $\Phi(\vartheta)$. The behavior of the function will depend on the explicit physical situation of the gas in general. However, the stability requirement, the inequality in Eq.~\eqref{stability:TTES}, highly constrains the functional form.
Rewriting by using Eqs.~\eqref{idgas1}, \eqref{rho:ig}, and \eqref{f'}, we get
\be{stb}
\frac{m/\Theta+ c_v +1}{c_v \Theta^2} \frac{\vartheta }{\Phi'(\vartheta)}  > 1.
\ee
This inequality constrains $\Phi'(\vartheta)$ should have the same signature as that of $\vartheta$.
Therefore, $\Phi(\vartheta)$ should be monotonically increases/decreases for $\vartheta \gtrless 0$.
In other words, there is no local minimum of $ \Phi(\vartheta)$ other than $\vartheta =0$.
The energy density, therefore, will have a minimum value at $\vartheta =0$ corresponding to the thermal equilibrium configuration. 
Consequently, $\Phi'(0)=0$ if the function $ \Phi(\vartheta)$ is a differentiable at $\vartheta =0$. 
This property is directly related to the stability of the thermal equilibrium system at $\vartheta=0$.
Now, the function $\Phi$ can be written in a series form,
\be{f:series}
 \Phi(\vartheta) = \frac{\gamma}2 \left(\frac{\vartheta}{\vartheta_0}\right)^2+ \frac{\gamma'}{3!} \left(\frac{\vartheta}{\vartheta_0}\right)^3+ \cdots .
\ee
Because $\vartheta =0$ is a minimum, $\gamma/\vartheta_0^2 > 0$.
Because of the property, $\Phi'(\vartheta)$ must be an invertible, one-to-one function around $\vartheta \sim 0$.
We write $\vartheta$ as a function of $Q$:
\be{vT:q}
\vartheta \equiv \vartheta(Q) = {\Phi'}^{-1}(\frac{q}{c_v n\Theta^2} ) 
	= \vartheta_0 Q
	\left[1+ \sum_{j=1}^\infty g_j Q^j \right], \qquad 
Q \equiv \frac{\vartheta_0 q}{\gamma c_v n\Theta^2} .
\ee
where the coefficient $g_j$ must be determined from the series form of $\Phi(\vartheta)$ in Eq.~\eqref{f:series}. 
To the linear order in $Q$, we have 
$$
\vartheta \approx \frac{\vartheta_0^2 q}{\gamma c_v n\Theta^2} .
$$

Note that, given $\vartheta$, the inequality~\eqref{stb} can be written as a third-order polynomial function for $1/\Theta$. 
By defining $X \equiv \sqrt{\vartheta/\Phi'(\vartheta)} /\Theta$, we rewrite 
the inequality~\eqref{stb} in the form:
\be{Theta >}
f(X) \equiv \frac{m}{(1+c_v)\Theta_{\rm ub}} X^3 + X^2 -1 > 0, \qquad \Theta_{\rm ub} \equiv
\sqrt{\frac{1+c_v}{ c_v} \frac{\vartheta}{\Phi'(\vartheta)}} .
\ee
Since mass is non-negative, $f(X)$ is a monotonically increasing function of $X > 0$.  
This constrains the value of $X$ to be larger than the root, $X_{\rm min} \geq 0$, of the equation, $f(X_{\rm min}) =0$. 
Ultimately, this inequality presents an upper bound on temperature in terms of given physical parameters:
\be{Theta max}
\Theta < \Theta_{\rm max}, \qquad \Theta_{\rm max}(\vartheta) = \frac{\Theta_{\rm ub}(\vartheta)}{X_{\rm min}},
\ee
where, $\Theta_{\rm max}$ denotes the upper bound.
The value of $X_{\rm min}$ reaches its maximum, $X_{\rm min}=1$, when the mass of the particles is negligible, i.e., $m \to 0$.
In this case, the upper bound on the temperature becomes
$
\Theta_{\rm max} = \Theta_{\rm ub} . 
$

Let us examine the result around the thermal equilibrium $q\sim 0$.
In this limit, $\lim_{q\to 0}\vartheta/\Phi'(\vartheta) = \vartheta_0^2/\gamma$, where the right-hand side is nothing but the coefficient of second order term in the series expansion of $\Phi(\vartheta)$.
Now, $\Theta_{\rm ub}$ becomes 
$$
\lim_{q\to 0} \Theta_{\rm ub} 
=\sqrt{\frac{1+c_v}{ c_v} \frac{\vartheta_0^2}{\gamma}}.
$$
It is noteworthy that the  maximum temperature tends to infinity only when $\gamma \to 0$, i.e., when the quadratic part of the series form of $\Phi(\vartheta)$ vanishes.
This observation is an interesting because once we understand how the energy density depends on heat, we can identify a maximum temperature for a stable equilibrium system based on physical parameters.

Let us calculate the function $\Phi(\vartheta)$ for the case of the monoatomic ideal gas in Eq.~\eqref{delta sigma} with $c_v = 3/2$.
Now, we compare this result with Eq.~\eqref{T:gen} to find:
\be{Phi:if}
\Phi(\vartheta) = \frac{5}{72} \left(\frac{q \delta L}{\kappa \Theta}\right)^2 .
\ee
To compare this form with Eq.~\eqref{f:series}, we write the heat conductivity $\kappa$ explicitly.
The heat conductivity of a gas is usually takes the form, 
$$
\kappa = \frac{3k_B}{2m} \varrho \frac{\Delta x^2}{\tau} 
= \frac{3k_B}{2m} n \langle \frac{m v^2}2\rangle \frac{2\tau}{3}  
=  \frac{3k_B}{4m} n  \Theta\tau
$$ 
where $\varrho$, $\Delta x$, $v\equiv \Delta x/\tau$, and $\tau$ are the mass density, the mean free path along the direction of heat flow, the velocity of particle between collisions, and the time between two nearby collisions of the gas particles (mean free time), respectively.

Substituting the result to Eq.~\eqref{Phi:if}, we obtain, for the gas system consisting of particles of mass $m>0$:
\be{ig 1}
\frac{\vartheta_0^2}{\gamma} 
= \frac{3(1+c_v)}{4 c_v^4} \left( \frac{ m \delta L}{\tau}\right)^2 
= \frac{10}{27} \left( \frac{  m \delta L}{\tau}\right)^2 
.
\ee
Therefore, $\Theta_{\rm ub}$ becomes
\be{Theta ub}
\Theta_{\rm ub} = \sqrt{\frac{1+c_v}{c_v} \frac{\vartheta_0^2}{\gamma}}
	=  \frac{5\sqrt{2}}{9 }  \left( \frac{ \delta L}{ c \tau}\right) \left(\frac{m c^2}{k_B}\right) .
\ee
In the last equality, we recover the light velocity $c$ and the Boltzmann constant $k_B$ to compare the value of the upper bound.
As seen in this equation, the upper bound is of the order of the mass-scale temperature times the ratio between the two length scale $\delta L$ and $c\tau$.
Here, the lowest value $\delta L$ can represent the scale that the ideal gas nature of the particles breaks down, such as the quantum de-Broglie length scale or the scale that the inter-particle interactions becomes important.
The term $c \tau$ represents the length that the light travels during the mean free time of the gas particles.
Note that we cannot use this result for a gas consisting of massless particles because the Newtonian heat equation and the formula for the average kinetic energy may not hold in the massless case.

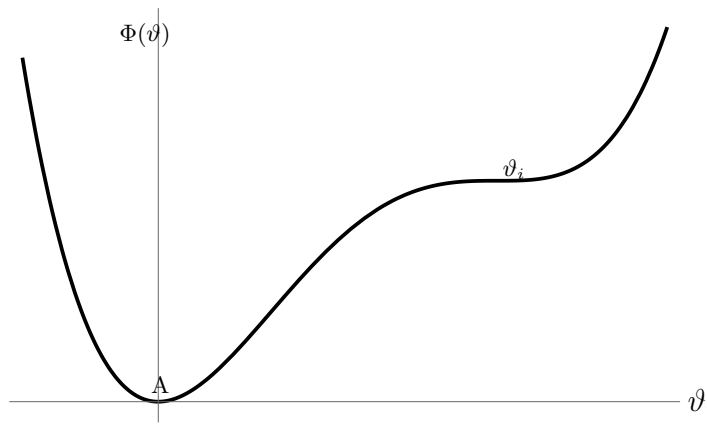
\begin{figure}[tbh]
\begin{center}
\begin{tabular}{c}
\includegraphics[width=.5\linewidth,origin=tl]{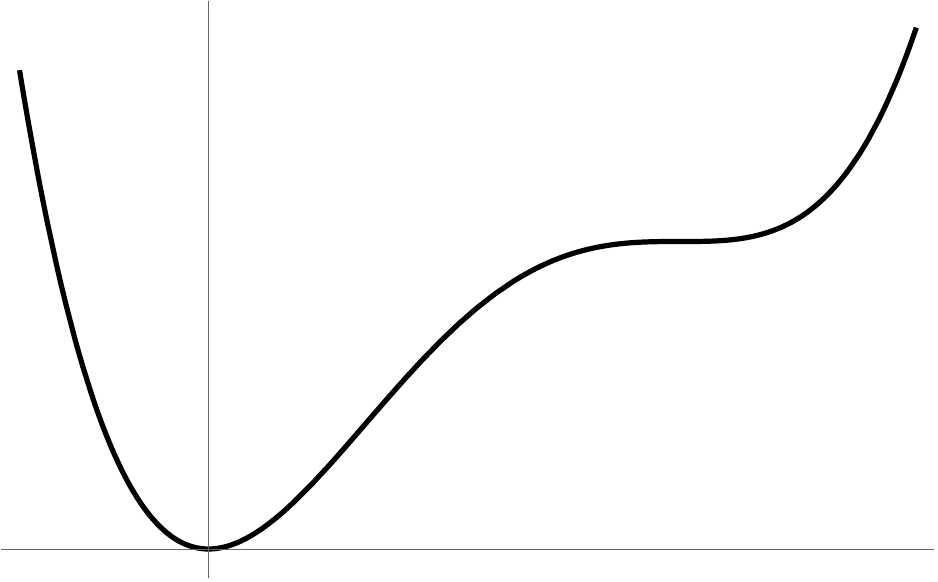} 
\end{tabular}
\put (0,-70){\large $\vartheta$}
%\put(-205,-50) {1}
\put (-70, 19) {$ \vartheta_i$}
\put(-203, -63){A}
\put(-215,70){$\Phi(\vartheta)$}
\end{center}
\caption{A general form for $\Phi(\vartheta)$. The function $ \Phi(\vartheta)$ has a global minimum `A' at $\vartheta =0$ and may have an inflection point $\vartheta_i$.}
\label{fig:vt}
\end{figure}

\begin{figure}[tbh]
\begin{center} 
\begin{tabular}{cc}
\includegraphics[width=.35\linewidth,origin=tl]{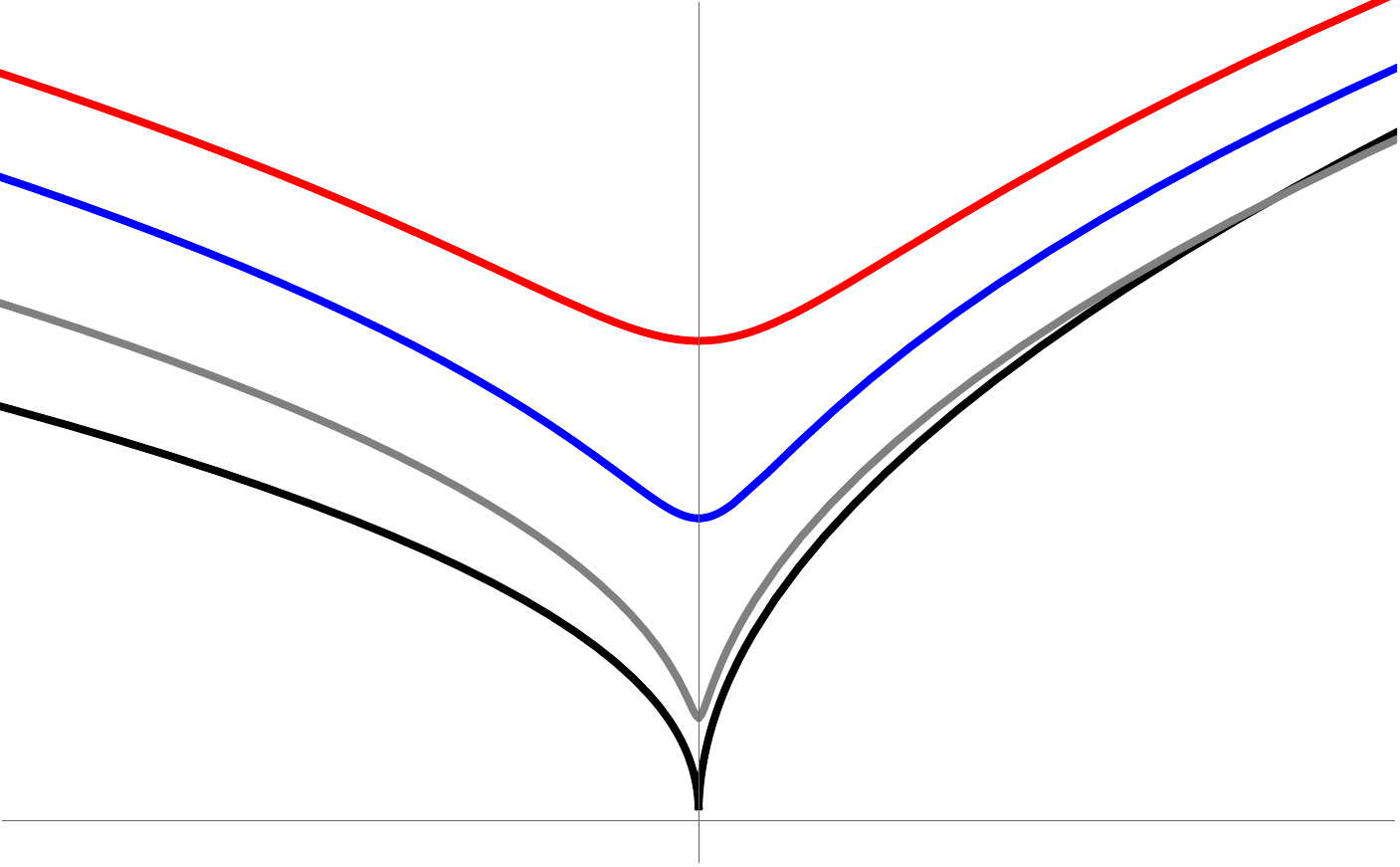}
&~~ \includegraphics[width=.35\linewidth,origin=tl]{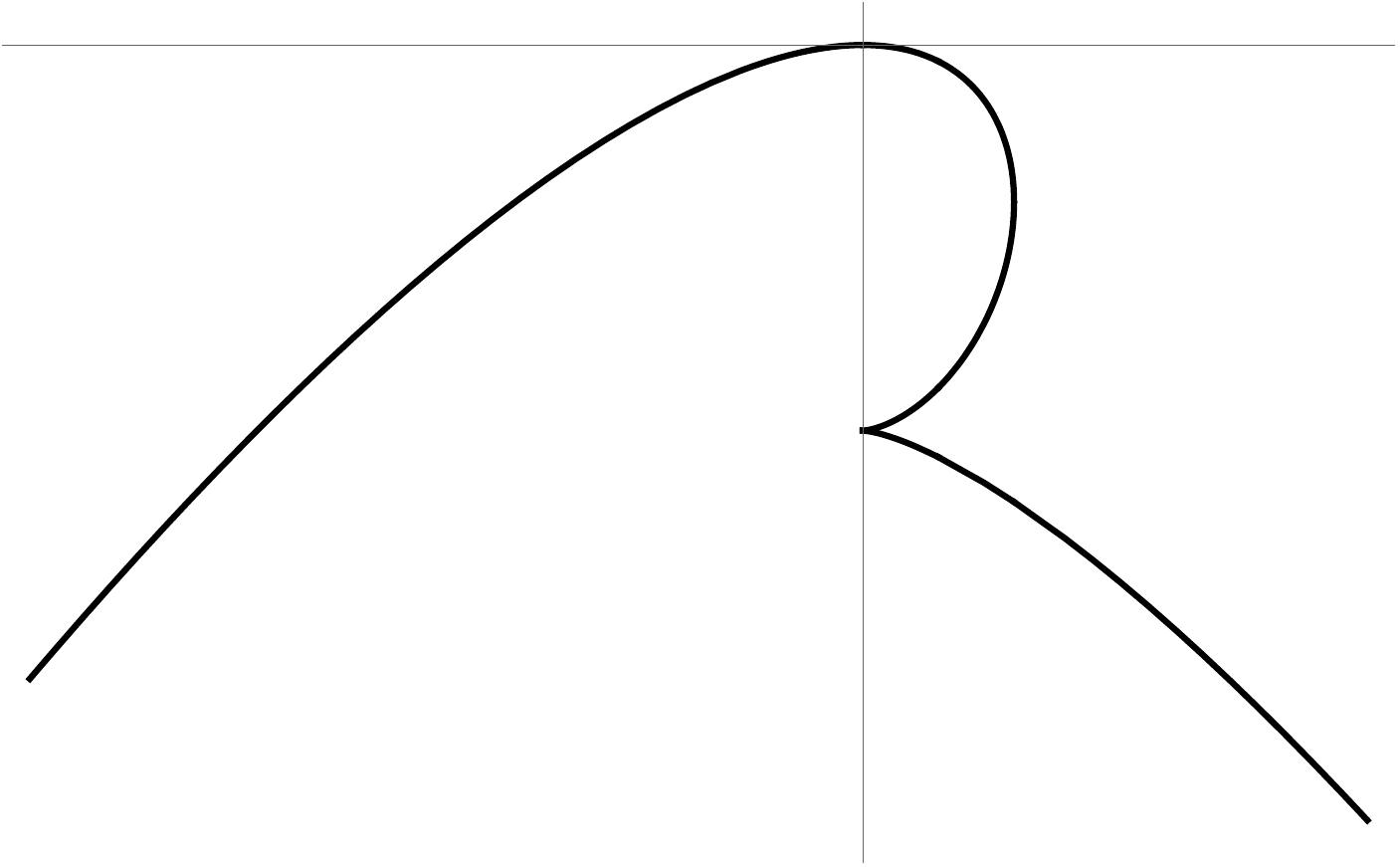}
\end{tabular}
\put (-5,50){ $ \frac{q}{c_v n \Theta^2}$}
%\put(-205,-50) {1}
\put (-90, 59)  { $ (\sigma-\sigma^*)/c_v $}
\put(-213, -43){$\frac{q}{c_v n_0 \Theta_0^2}$}
\put(-80,-0){$\vartheta_i$}
\put(-305,60){$\Theta/\Theta_0$}
\end{center}
\caption{The behaviors of the temperature (L) and the specific entropy (R) with respect to the change of  heat $q$ for the behavior $\Phi(\vartheta)$ in Fig.~\ref{fig:vt}.
In the left panel, we choose $c_v =3/2$, $\sigma =1$, and $n/n_0 = 2, 1, 1/5, $ and $ 1/100$, respectively from the top.  
The temperature and the specific entropy take its minimum/maximum at $q=0$.
Here, $\sigma^*$ is given in Eq.~\eqref{sigma*}.
The minimum/maximum value is nothing but the thermal equilibrium ones. 
}
\label{fig:Tsigma}
\end{figure}
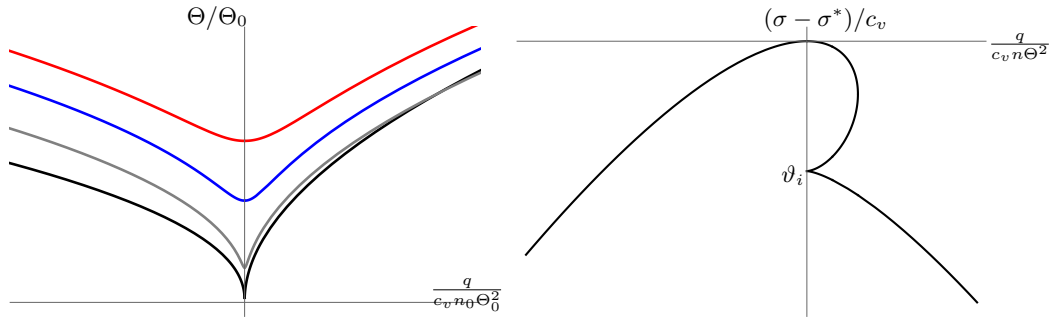

\vspace{.3cm}
Let us describe the behaviors of the temperature, entropy, internal energy and physical behaviors for a generic form of $\Phi(\vartheta)$. 
As an example of generic form~\eqref{f:series}, we consider $\Phi(\vartheta)$ shown in Fig.~\ref{fig:vt}, which has an inflection point at $\vartheta_i$ with $\Phi'(\vartheta_i) =0$.
Now,  from $\Phi'(\vartheta) =0$ in Eq.~\eqref{f'}, the absence of heat, $q=0$, corresponds to two distinct values, the minimum, $\vartheta =0$, and the inflection point, $\vartheta =\vartheta_i$.
Note that heat flows along the same direction at both sides of the inflection point.
The inequality in Eq.~\eqref{stb} still holds.
Around the inflection point, the function takes the form, 
$
\Phi \approx \Phi(\vartheta_i) + \alpha/3!\times (\vartheta -\vartheta_i)^3 +\cdots
$
 with 
$\alpha \gtrless 0$ for $\vartheta_i \gtrless 0$.
Then, $|q|/n\Theta^2 \approx (\alpha c_v/2) (\vartheta- \vartheta_i)^2$ increases as $\vartheta$ becomes farther from the inflection point.
This behavior is evident in the right-panel of Fig.~\ref{fig:Tsigma}.
Note also that the maximum temperature at $\vartheta_i$ is infinite.
The energy density $\rho(n,s,\vartheta_i)$ is higher than that at $\vartheta =0$ because $\Phi(\vartheta_i) > \Phi(0)$.
This energy difference will infer a stability issue for the system in the inflection point and suggests a kind of phase transition from one state $\vartheta = \vartheta_i$ to the equilibrium state at $\vartheta=0$.

Let us illustrate the generic behaviors of the temperature, entropy, internal energy as we vary $q$. 
From Eqs.~\eqref{T:gen} and \eqref{f'}, we can express heat as a function of $n$, $s$, and $\Theta$: 
\be{q:T}
q = c_v n_0 \Theta_0^2 \frac{n}{n_0} \left(\frac{\Theta}{\Theta_0}\right)^2 \Phi' \left( \Phi^{(-1)} \Big(
\log \frac{\Theta}{\Theta_0} - \frac{\sigma +\log n/n_0}{c_v}\Big) \right) 
\ee
where $\Phi^{(-1)}$ denotes the inverse function of $\Phi(\vartheta)$.
We interpret this equation as an implicit function between the temperature $\Theta$ and the heat $q$. 
We then plot the temperature as a function of $q$ in the left panel of Fig.~\ref{fig:Tsigma} for various $n$ and $s$ for $\Phi(\vartheta)$ in Fig.~\ref{fig:vt}.
A crucial observation is that temperature reaches its minimum value at $q=0$ and increases with $|q|$.
We can also utilize this equation to express the specific entropy as a function of $n$, $\Theta$, and $q$: 
\be{sigma:q}
\sigma(n, \Theta, q) - \sigma^*(n,\Theta) = - c_v \Phi\left( (\Phi')^{(-1)}\big(
	\frac{q}{c_v n \Theta^2}\big) \right) .
\ee
Here $\sigma^*$ is defined in Eq.~\eqref{sigma*}.
The contribution of heat to the specific entropy is plotted in the right panel of Fig.~\ref{fig:Tsigma}.
As shown here, the (specific) entropy is maximized when $q=0$ and $\vartheta =0$. 
At the inflection point, $\vartheta_i$, the heat vanishes but the entropy does not have a local maximum.  
The behavior of the internal energy $\rho$ and kinetic energy with respect to $q$ are determined implicitly through the dependence of $\Theta$ on $q$ because of Eq.~\eqref{rho:ig}.  

\vspace{.3cm}
When energy is poured into an ideal gas system while holding $\vartheta$ and $n$ constant, entropy increases indefinitely as dictated by the first law of thermodynamics~\eqref{1st law}.
Subsequently, Eq.~\eqref{T:gen} appears to show that the temperature also increases without bound.
On the contrary, beyond the `Hagedorn temperature'~\cite{Hagedorn} for hadrons, the particle number cannot remain constant but instead increases with energy, while keeping the temperature bounded. 
Assuming a Hagedorn-like temperature, $T_{\rm Hagedorn}$, to be the temperature upper-bound for an ideal gas, we  determine the quadratic part of the $\Phi$~\eqref{f:series} to the form:
$$
\frac{\gamma} {\vartheta_0^2}= \frac{1+c_v}{c_v  T_{\rm Hagedorn}^2} \quad \rightarrow \quad
\Phi(\vartheta) \approx \frac{1+c_v}{2c_v}\frac{\vartheta^2}{ T_{\rm Hagedorn}^2 } .
$$
This equation tells that the heat contribution to the energy density is suppressed quadratically by the Hagedorn temperature.
Consequently, in the absence of the bound (when the bound diverges), the energy density should be independent of heat. 

The temperature upper bound established in this work stems from the stability condition of a thermal equilibrium state in relativistic thermodynamics.
This condition is generalized to a relation for a boost transformation between the Eckart and Landau-Lifshitz observers, rooted in the theory of relativity.
Since the derivation is grounded in first principles, the result may not be limited to the ideal gas but could be universal for various forms of matter.
In the present formulation, the upper bound of the temperature depends on the value $\gamma/\vartheta_0^2$, which must be independent of other thermodynamic parameters by definition.
For the case of a simple ideal gas, the upper bound is of the order of the mass-scale temperature times the ratio between the two length scale $\delta L$ and $c\tau$~\eqref{Theta ub}.
The undetermined length $\delta L$ must be short enough in a macroscopic scale but large enough in a microscopic scale. 
The mean free time $\tau$ of the particles is dependent on the microphysics of the theory. 
In a general case, the mean free time may depend on the temperature and the number density of particles.
In that case, a better description for the gas will be required. 
% ideal gas assumption may fail.

%=================================================
\section*{Acknowledgment}
This work was supported by the National Research Foundation of Korea grants funded by the Korea government RS-2023-00208047.
The author thanks to APCTP and CQUeST for hospitality.
The author also thanks to Prof. Jungjai Lee and Prof. Jaeweon Lee for helpful discussions.
%\appendix

%======================================

\end{document}